\documentclass[fleqn,10pt]{wlscirep}
\usepackage{amsmath, alltt, amssymb, xspace, times, epsfig,algorithm, algpseudocode}
\usepackage{amsthm, amsfonts, latexsym, graphicx, color, hyperref}
\usepackage{mathtools}
\usepackage{tikz}
\usepackage{xr}
\theoremstyle{plain}
\newtheorem{theorem}{Theorem}
\newtheorem{lemma}{Lemma}
\newtheorem{claim}{Claim}

\newtheorem{definition}{Definition}

\newcommand{\ii}{\mathbb{I}}

\newcommand{\ket}[1]{\vert #1 \rangle}

\newlength{\actualtopmargin}
\newlength{\actualsidemargin}
\title{Solving Set Cover with Pairs Problem Using Quantum Annealing}

\author[1,*]{Yudong Cao}
\author[1,$\dagger$]{Shuxian Jiang}
\author[2]{Debbie Perouli}
\author[3,4]{Sabre Kais}
\affil[1]{Department of Computer Science, Purdue University, West Lafayette, IN 47906, USA}
\affil[2]{Department of Mathematics, Statistics and Computer Science, Marquette University, Milwaukee, WI 53233, USA}
\affil[3]{Department of Chemistry, Physics and Birck Nanotechnology Center, Purdue University, West Lafayette, IN 47906, USA}
\affil[4]{Qatar Energy and Environment Research Institute (QEERI), HBKU, Doha, Qatar}

\affil[*]{cao23@purdue.edu, $\dagger$ jiang97@purdue.edu}

\begin{abstract}

Here we consider using quantum annealing to solve Set Cover with Pairs (SCP), an NP-hard combinatorial optimization problem that play an important role in networking, computational biology, and biochemistry. We show an explicit construction of Ising Hamiltonians whose ground states encode the solution of SCP instances. We numerically simulate the time-dependent Schr\"{o}dinger equation in order to test the performance of quantum annealing for random instances and compare with that of simulated annealing. We also discuss explicit embedding strategies for realizing our Hamiltonian construction on the D-wave type restricted Ising Hamiltonian based on Chimera graphs. Our embedding on the Chimera graph preserves the structure of the original SCP instance and in particular, the embedding for general complete bipartite graphs and logical disjunctions may be of broader use than that the specific problem we deal with.

\end{abstract}
\begin{document}

\flushbottom
\maketitle
%
%
\thispagestyle{empty}


\section{Introduction}


Quantum annealing (QA) uses the principles of quantum mechanics for solving unconstrained optimization problems\cite{FGSS+94,kadowaki1998quantum,farhi2001quantum,DC05}. Since the initial proposal of QA, there has been much interest in the search for practical problems where it can be advantageous with respect to classical algorithms\cite{DC05, DC08, HJLB+10, JAG+11, BFKS+13, DJAH+13, BRIW+14, MW13, Dash13, BASC+13, LPSS+14, SQVL14, RWJB+14, VMBR+14, SSSV14, AVMW+15, G14, VMKO15, VAPI+15, ARTL15, KM14, CDVW+14, HJAR+15, SRT15, BWPT15, AHSL15, KHZO+15, CSVA+16, POFB+15, VAL15}, particularly simulated annealing (SA)\cite{FGG02,SMTC02,HRIT14}. Extensive theoretical, numerical and expeirmental efforts have been dedicated to studying the performance of quantum annealing on problems such as satisfiability\cite{FGGS00,FGG00SAT,choi2010adiabatic}, exact cover\cite{farhi2001quantum,choi2010adiabatic}, max independent set\cite{choi2010adiabatic}, max clique\cite{CFGG02}, integer factorization\cite{peng2008quantum}, graph isomorphism~\cite{hen2012solving, gaitan2014graph}, ramsey number\cite{bian2013experimental}, binary classification\cite{neven2008training,denchev2012robust}, unstructured search\cite{RC02} and search engine ranking\cite{garnerone2012adiabatic}. Many of these approaches\cite{FGGS00,farhi2001quantum,peng2008quantum,hen2012solving,FGG00SAT,CFGG02, gaitan2014graph,bian2013experimental,neven2008training,denchev2012robust} recast the computational problem at hand into a problem of finding the ground state of a quantum Ising spin glass model, which is NP-complete to solve in the worst case\cite{Barahona82,lucas2013ising}. 

The computational difficulty of Ising spin glass has not only given the quantum Ising Hamiltonians the versatility for efficiently encoding many problems in NP\cite{lucas2013ising}, but also motivated physical realization of QA using systems described by the quantum Ising model\cite{HJLB+10,JAG+11,DJAH+13}. The notion of adiabatic quantum computing (AQC)\cite{farhi2001quantum,crosson2014different,FGGS00}, which can be regarded as a particular class of QA, has further established QA in the context of quantum computation ({In this work we will use the terms quantum annealing and adiabatic quantum computing synonymously}). Although it is believed that even universal quantum computers cannot solve \textsc{NP}-complete problems efficiently in general \cite{Aaronson10}, there has been evidence in experimental quantum Ising systems that suggests quantum speedup over classical computation due to quantum tunneling \cite{NSK12,denchev2015computational}. It is then of great interest to explore more regimes where quantum annealing could offer a speedup compared with simulated annealing.

Here we consider a variant of Set Cover (SC) called Set Cover with Pairs (SCP). SC is one of Karp's 21 NP-complete problems\cite{Karp1972} and SCP was first introduced\cite{HS05} as a generalization of SC. Instead of requiring each element to be covered by a single object as in SC, the SCP problem is to find a minimum subset of objects so that each element is covered by \emph{at least one pair} of objects. We will present its formal definition in Section \ref{sec:prelim}. SCP and its variants arise in a wide variety of contexts including Internet traffic monitoring and content distribution\cite{BDD+11}, computational biology\cite{LPR04,WY11}, and biochemistry\cite{gonccalves2012exact}.
On classical computers, the SCP problem is at least as hard to approximate as SC. Specifically, its difficulty on classical computers can be manifested in the results by Breslau {\em et al}\cite{BDD+11}, which showed that no polynomial time algorithm can approximately solve Disjoint-Path Facility Location, a special case of SCP, on $n$ objects to within a factor that is $2^{\log^{1-\epsilon}{n}}$ for any $\epsilon>0$. Due to its complexity, various heuristics\cite{HS05} and local search algorithms\cite{gonccalves2012exact} have been proposed. 

In this paper we explore using quantum annealing based on Ising spin glass to solve SCP. We start by reducing SCP to finding the ground state of Ising spin glass, via integer linear programming (Theorem \ref{thm:scp_ising}). We then simulate the adiabatic evolution of the time dependent transverse Ising Hamiltonian $H(s)=(1-s)H_0+sH_1$ which interpolates linearly between an initial Hamiltonian $H_0$ of independent spins in uniform transverse field and a final Hamiltonian $H_1$ that encodes an SCP instance. For randomly generated SCP instances that lead to Ising Hamiltonian constructions of up to 19 spins, we explicitly simulate the time dependent Schr\"{o}dinger equation. We compute the minimum evolution time that each instance needed to accomplish 25\% success probability. For benchmark purpose we also use simulate annealing to solve the instances and compare its performance with that of adiabatic evolution. Results show that the median time for yielding 25\% success probablity scales as $O(2^{0.33M})$ for quantum annealing and $O(2^{0.21M})$ for simulated annealing, observing no general quantum speedup. However, the performance of quantum annealing appears to have wider range of variance from instance to instance than simulated annealing, casting hope that perhaps certain subsets of the instance could yield a quantum advantage over the classical algorithms.

Aside from the theoretical and numerical studies, we also consider the potential implementation our Hamiltonian construction on the large-scale Ising spin systems manufactured by D-Wave Systems\cite{JAG+11,HJLB+10,DJAH+13,LPSS+14}. Benchmarking the efficiency of QA is currently of significant interest. An important issue that needs to be addressed in such benchmarks is that the physical implementation of the algorithm could be affected by instance-specific features. This is manifested in the embedding\cite{KSH12,choi11} of the Ising Hamiltonian construction onto the specific topology of the hardware (the Chimera graph\cite{bian2014discrete,KSH12,VMKO15}). Here we present a general embedding of SCP instances onto a Chimera graph that preserves the original structure of the instances and requires less qubits than the usual approach by complete graph embedding. This allows for efficient physical implementations that are untainted by ad hoc constructions that are specific to individual instances. 

\section{Preliminaries}\label{sec:prelim}

\subsection{Set Cover with Pairs}\label{subsec:scp}

Given a \emph{ground set} $U$ and a collection $S$ of subsets of $U$, which we call the \emph{cover set}. Each element in $S$ has a non-negative weight, the {Set Cover} (SC) problem asks to find a minimum weight subset of $S$ that covers all elements in $U$. Define \emph{cover function} as $Q:S\mapsto 2^U$ where $\forall s\in S$, $Q(s)$ is the set of all elements in $U$ covered by $s$. Then SC can be formulated as finding a minimum weight $S'\subseteq S$ such that $Q(S')=\cup_{s'\in S'}Q(s')=U$. Set Cover with Pairs (SCP) can be considered as a generalization of SC in the sense that if we define the cover function such that $\forall i,j\in S$, $i\neq j$, $Q(i,j)$ is the set of elements in $U$ covered by the pair $\{i,j\}$, then SCP asks to find a minimum subset $A\subseteq S$ such that $Q(A)=\cup_{\{i,j\}\in S}Q(i,j)=U$. Here we restrict to cases where each element of $S$ has unit weight.

A \emph{graph} $G(V,E)$ is a set of vertices $V$ connected by a set of edges $E$. A \emph{bipartite graph} is defined as a graph  whose set of vertices $V$ can be partitioned into two disjoint sets $V_1$ and $V_2$ such that no two vertices within the same set are adjacent. We formally define SCP as the following.

\begin{definition}\label{def:SCPP}
	({Set Cover with Pairs}) Let $U$ and $S$ be disjoint sets of elements and $V=U\cup S$. Given a bipartite graph $G(V,E)$ between $U$ and $S$ with $E$ being the set of all edges, find a subset $A\subseteq S$ such that:
	\begin{enumerate}
		\item $\forall c_i\in U$, $\exists a_1^{(i)},a_2^{(i)}\in A$ such that $(a_1^{(i)},c_i)\in E$ and $(a_2^{(i)},c_i)\in E$. In other words, $c_i$ is covered by the pair $\{a_1^{(i)},a_2^{(i)}\}$.
		\item The size of the set, $|A|$, is minimized.
	\end{enumerate}
	We use the notation $\textsc{SCP}(G,U,S)$ to refer to a problem instance with $|U|=n$, $|S|=m$ and the connectivity between $U$ and $S$ determined by $G$.
\end{definition}

\subsection{Quantum annealing, adiabatic quantum computing}

In this paper we use QA as a heuristic method to solve the SCP problem. QA was proposed~\cite{kadowaki1998quantum} for solving optimization problems using quantum fluctuations, known as quantum tunneling, to escape local minima and discover the lowest energy state. Farhi {\em et al.}~\cite{farhi2001quantum} provide the framework for using Adiabatic Quantum Computation (AQC), which is closely related to QA, as a quantum paradigm to solve NP-hard optimization problems. The first step of the framework is to define a Hamiltonian $H_P$ whose ground state corresponds to the solution of the combinatorial optimization problem. Then, we initialize a system in the ground state of some beginning Hamiltonian $H_B$ that is easy to solve, and perform the adiabatic evolution $H(s)=(1-s)H_B+sH_P$. Here $s\in[0,1]$ is a time parameter. In this paper we only consider time-dependent function $s(t)=t/T$  for total evolution time $T$, but in general it could be any general functions that satisfy $s(0)=0$ and $s(T)=1$. The adiabatic evolution is governed by the Schr\"{o}dinger equation
\begin{equation}\label{eq:Schr}
i\frac{d}{dt}\ket{\psi(t)}=H(s(t))\ket{\psi(t)}
\end{equation}
where $|\psi(t)\rangle$ is the state of the system at any time $t\in[0,T]$. Let $\pi_i(s)$ be the $i$-th instantaneous eigenstate of $H(s)$. In other words, let $H(s)\ket{\pi_i(s)}=E_i(s)\ket{\pi_i(s)}$ for any $s$. In particular, let $|\pi_0(s)\rangle$ be the instantaneous ground state of $H(s)$.

According to the adiabatic theorem\cite{M62}, for $s$ varying sufficiently slow from 0 to 1, the state of the system $|\psi(t)\rangle$ will remain close to the true ground state $|\pi_0(s(t))\rangle$. At the end of the evolution the system is roughly in the ground state of $H_P$, which encodes the optimal solution to the problem. If the ground state of $H_P$ is NP-complete to find (for instance consider the case for Ising spin glass\cite{Barahona82}), then the adiabatic evolution $H(s)$ could be used as a heuristic for solving the problem.

An important issue associated with AQC is that the adiabatic evolution needs to be slow enough to avoid exciting the system out of its ground state at any point. In order to estimate the scaling of the minimum runtime $T$ needed for the adiabatic computation, criteria based on the minimum gap between the ground state and the first excited state of $H(s)$ is often used. However, here we do not use the minimum gap as an intermediate for estimating the runtime scaling, but instead numerically integrate the time dependent Schr\"{o}dinger equation \eqref{eq:Schr}.

\subsection{Quantum Ising model with transverse field}

The Hamiltonian for an Ising spin glass on $N$ spins can be written as
\begin{equation}\label{eq:ising}
H=\sum_{i=1}^N h_i\sigma_i^z+\sum_{i<j}^NJ_{ij}\sigma_i^z\sigma_j^z
\end{equation}
where $\sigma_i^z={\ii}^{\otimes(i-1)}\otimes\left(\begin{smallmatrix}1 & 0 \\ 0 & -1\end{smallmatrix}\right)\otimes{\ii}^{\otimes(n-i)}$ acts on the $i$-th spin with ${\ii}$ being a $2\times 2$ identity matrix. $h_i$, $J_{ij}$ are coefficients. 
The Hamiltonian is diagonal in the basis $\{|{\bf s}\rangle\in\mathbb{C}^{2^N}|{\bf s}\in\{0,1\}^N\}$ in the Hilbert space $\mathcal{H}$. In particular $\sigma^z|0\rangle=|0\rangle$ and $\sigma^z|1\rangle=-|1\rangle$.
We formally define the problem of finding the ground state of an $N$-qubit Ising Hamiltonian in the following.
\begin{definition}\label{def:ising}
	({Ising Hamiltonian}) Given the Hamiltonian $H$ in equation \eqref{eq:ising}, find a quantum state $|{\bf s}\rangle\in\mathcal{H}$ , where $\mathcal{H}$ is $2^N$-dimensional, such that the energy  $E({\bf s})=\langle{\bf s}|H|{\bf s}\rangle$ is minimized. We use the notation \textsc{Ising}$({\bf h},{\bf J})$ to refer to the problem instance where ${\bf h}=(h_1,h_2,\cdots,h_N)^T$ and ${\bf J}\in\mathbb{R}^{N\times N}$ is a matrix such that the $ij$-th and the $ji$-th elements are equal to $J_{ij}/2$. The diagonal elements of $\bf J$ are 0. Hence $E({\bf s})={\bf h}^T{{\bf p}(\bf s)}+{\bf p}({\bf s})^T{\bf J}{\bf p}({\bf s})$ where ${\bf p}({\bf s})=1-2{\bf s}\in\{-1,1\}^N$.
\end{definition}

In this paper, we construct Ising Hamiltonians whose ground state encodes the solution to an arbitrary instance of the SCP problem. The physical system used for quantum annealing that we assume is identical to that of D-Wave\cite{JAG+11,HJLB+10,DJAH+13,LPSS+14}, namely Ising spin glass with transverse field
\begin{equation}\label{eq:tising}
H=\sum_{i=1}^N\Delta_i\sigma_i^x+\sum_{i=1}^N h_i\sigma_i^z+\sum_{i<j}^NJ_{ij}\sigma_i^z\sigma_j^z
\end{equation}
where $\sigma_i^x={\ii}^{\otimes(i-1)}\otimes\left(\begin{smallmatrix}0 & 1 \\ 1 & 0\end{smallmatrix}\right)\otimes{\ii}^{\otimes(n-i)}$ acts on the $i$-th spin. The beginning Hamiltonian $H_B$ has its $h_i,J_{ij}=0$ for all $i,j$ and the final Hamiltonian $H_P$ has $\Delta_i=0$ for all $i$ while $h_i$ and $J_{ij}$ depend on the problem instance at hand. We will elaborate on assigning $h_i$ and $J_{ij}$ coefficients in $H_P$ in Theorem \ref{thm:scp_ising}.

\subsection{Graph minor embedding}\label{subsec:graph_minor}

The interactions described by the transverse Ising Hamiltonian in equation \eqref{eq:tising} are not restricted by any constrains. However, in practice the topology of interactions is always constrained to the connectivity that the hardware permits. Therefore in order to physically implement an arbitrary transverse Ising Hamiltonian, one must address the problem of embedding the Hamiltonian into the logical fabric of the hardware\cite{KSH12,choi11}. For convenience we define the \emph{interaction graph} of an Ising Hamiltonian $H$ of the form in equation \eqref{eq:ising} as a graph $G_H(V_H,E_H)$ such that each spin $i$ maps to a distinctive element $v_i$ in $V_H$ and there is an edge between $v_i$ and $v_j$ iff $J_{ij}\neq 0$. This definition also applies to the transverse Ising system described in equation \eqref{eq:tising}. We use the term \emph{hardware graph} to refer to a graph whose vertices represent the qubits  in the hardware and the edges describe the allowed set of couplings in the hardware.

In Section \ref{subsec:scp} we defined bipartite graphs. Here we define a \emph{complete bipartite graph $K_{m,n}$} as a bipartite graph where $|V_1|=m$, $|V_2|=n$ and each vertex in $V_1$ is connected with each vertex in $V_2$. A graph $H(W,F)$ is a \emph{subgraph} of $G(V,E)$ if $W\subseteq V$ and $F\subseteq E$. It is possible that the interaction graph of the desired Ising Hamiltonian is a subgraph of the hardware connectivity graph. In this case the embedding problem can be solved by \emph{subgraph embedding}, which we define as the following.

\begin{definition}\label{def:subgraph}
A \emph{subgraph embedding} of $G(V,E)$ into $G'(V',E')$ is a mapping $f:V\mapsto V'$ such that each vertex in $V$ is mapped to a unique vertex in $V'$ and if $(u,v)\in E$ then $(f(u),f(v))\in E'$.
\end{definition}

In more general cases, for an arbitrary Ising Hamiltonian, a subgraph embedding may not be obtainable and we will need to embed the interaction graph into the hardware as a graph \emph{minor}. Before we define minor embedding rigorously, recall that a graph is \emph{connected} if for any pair of vertices $u$ and $v$ there is a path from $u$ to $v$. A \emph{tree} is a connected graph which does not contain any simple cycles as subgraphs. $T$ is a \emph{subtree} of $G$ if $T$ is a subgraph of $G$ and $T$ is a tree. We then define minor embedding as the following.

\begin{definition}\label{def:minor}
A \emph{minor embedding} of $G(V,E)$ in $G'(V',E')$ is defined by a mapping $\phi:V\mapsto V'$ such that each vertex $v\in V$ is mapped to a connected subtree $T_v$ of $G'$ and if $(u,v)\in E$ then there exist $i_u,i_v\in V'$ such that $i_u\in T_u$, $i_v\in T_v$ and $(i_u,i_v)\in E'$.
\end{definition}

If such a mapping $\phi$ exists between $G$ and $G'$, we say $G$ is a \emph{minor} of $G'$ and we use $G\le_m G'$ to denote such relationship. Our goal is to take the interaction graph $G_{H}$ of our Ising Hamiltonian construction and construct the mapping $\phi$ that embeds $G_H$ into the hardware graph as a minor.

\subsection{Chimera graphs}\label{subsec:chimera}

\begin{figure}
\hspace*{-0.5cm}
\includegraphics[scale=0.8]{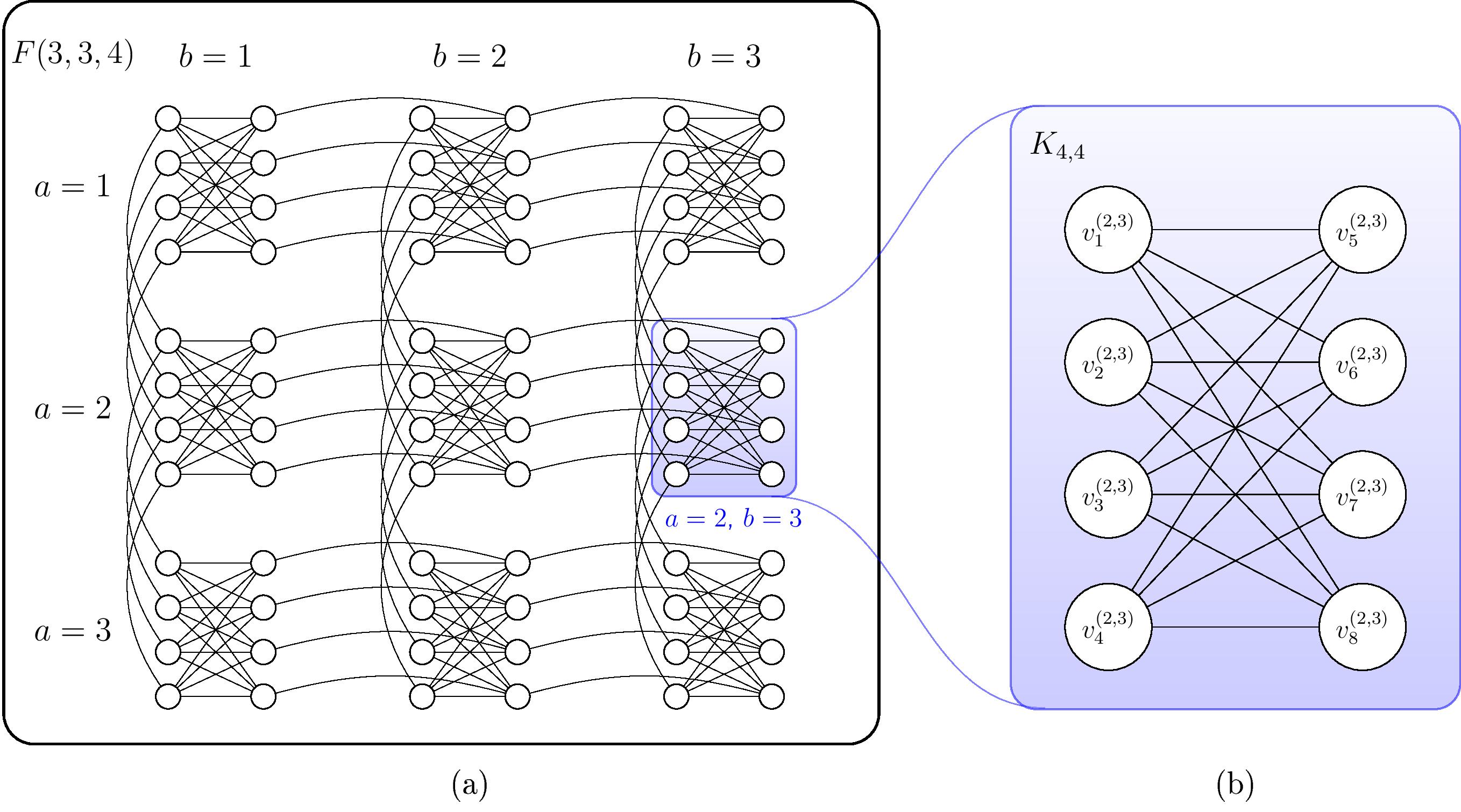}
\caption{The Chimera graph that represents the qubit connectivity of D-Wave hardware. (a) Example of a $3\times 3$ grid of $K_{4,4}$ cells, denoted as $F(3,3,4)$. (b) Labelling of nodes within a particular cell on the $a$-th row and $b$-th column. Here we use the cell on the 2nd row and 3rd column as an example.}
\label{fig:chimera_notation}
\end{figure}

Here we specifically consider the embedding our construction into a particular type of hardware graphs used by D-Wave devices\cite{bian2013experimental,ODD+12} called the \emph{Chimera graphs}. The basic components of this graph are 8-spin unit cells\cite{HJLB+10} whose interactions form a $K_{4,4}$. The $K_{4,4}$ unit cells are tiled together and the 4 nodes on the left half of $K_{4,4}$ are connected to their counterparts in the cells above and below. The 4 nodes on the right half of $K_{4,4}$ are connected to their counterparts in the cells left and right. Furthermore, we define $F(p,q,c)$ as a Chimera graph formed by an $p\times q$ grid of $K_{c,c}$ cells. Figure \ref{fig:chimera_notation}a shows $F(3,4)$ as an example. Note that any $K_{m,n}$ with $m,n\le c$ can be trivially embedded in $F(p,q,c)$ with any $p,q\ge 1$ via subgraph embedding. However, it is not clear \emph{a priori} how to embed $K_{m,n}$ with $m>c$ or $n>c$ onto a Chimera graph, other than using the general embedding of an $(m+n)$-node complete graph and consider $K_{m,n}$ as a subgraph. This costs $O((m+n)^2)$ qubits in general and one may lose the intuitive structure of a bipartite graph in the embedding. One of the building blocks of our embedding for our Ising Hamiltonian construction (Section \ref{subsec:embed_chimera}) is an alternative embedding strategy for mapping any $K_{m,n}$ onto $F(\lceil n/c\rceil,\lceil m/c\rceil,c)$ as a graph minor. Our embedding costs $O(mn)$ qubits and preserves the structure of the bipartite graph.

\section{Quantum annealing for solving SCP}

\subsection{From an arbitrary SCP instance to an Ising Hamiltonian construction}

SCP is NP-complete most simply because Set Cover (SC) is a special case of SCP\cite{HS05} and a solution to SCP is clearly efficiently verifiable. Since SC is NP-complete itself, any SCP instance can be rewritten as an instance of SC with polynomial overhead. The Ising Hamiltonian construction for Set Cover is explicitly known\cite{choi2010adiabatic,lucas2013ising}. Hence it is natural to consider using the chain of reductions from SCP to SC and then from SC to \textsc{Ising} (Definition \ref{def:ising}). If we recast each \text{SCP}$(G,U,S)$ with $|S|=m$ into an SC instance with a cover set of size $O(m^2)$. Using the construction by Lucas\cite{lucas2013ising} we have an Ising Hamiltonian
\begin{equation}\label{eq:ising_lucas}
H = H_A+H_B=A\sum_{\alpha=1}^n(1-\sum_{i:\alpha\in V_i}x_i)^2+B\sum_{i=1}^Nx_i
\end{equation}
where $V_i$ is the $i$-th cover set in the SC instance. Since the cover set $\{V_i\}$ is possibly of size up to $O(m^2)$, this leads to the Ising Hamiltonian in equation \eqref{eq:ising_lucas} costing $O(nm^2)$ qubits.

Here we present an alternative Ising Hamiltonian construction for encoding the solution to any SCP instance. We state the result precisely as Theorem \ref{thm:scp_ising} below. The qubit cost of our construction is comparable to that of Lucas. However, in Section \ref{subsec:embed_chimera} we argue that our construction affords more advantages in terms of embedding.

\begin{theorem}\label{thm:scp_ising}
	Given an instance of the Set Cover with Pairs Problem $\textsc{SCP}(G,U,S)$ as in Definition \ref{def:SCPP}, there exists an efficient (classical) algorithm that computes an instance of the Ising Hamiltonian ground state problem $\textsc{Ising}({\bf h},{\bf J})$ with ${\bf h}\in\mathbb{R}^{M}$ and ${\bf J}\in\mathbb{R}^{M\times M}$ where the number of qubits involved in the Hamiltonian is $M=O(nm^2)$ with $n=|U|$ and $m=|S|$.
\end{theorem}

\noindent{\emph{Proof.}} First, we recast an \textsc{SCP} instance to an instance of integer programming, which is \textsc{NP}-hard in the worst case. Then, we convert the integer programming problem to an instance of the \textsc{Ising} problem. Recall Definition~\ref{def:SCPP} of an \textsc{SCP}$(G,U,S)$ instance, where $G(V,E)$ is a graph on the vertices $V=U\cup S$. For each pair $f_i,f_j\in S$ define a set $Q_{ij}=\{c_k\in U|(f_i,c_k)\in E$ and $(f_j,c_k)\in E\}$. The problem can be recast as an integer program by

\begin{alignat}{2}
\min\quad & \sum_{f_i\in S}s_i &{}& \tag{LP} \label{ILP}\\
\mbox{s.t.}\quad &\sum_{c_k\in Q_{ij}}t_{ij}\ge 1 &\quad& \forall c_k\in U \tag{LP.1} \label{ILP-constr1}\\
& t_{ij}\le s_i \text{ and } t_{ij}\le s_j &{}& \forall f_i\neq f_j, \text{where } f_i, f_j\in S \text{ and } i<j \tag{LP.2}\label{ILP-constr2} \\
& s_i, t_{ij}\in\{0,1\} &{}& \forall f_i\neq f_j, \text{where } f_i,f_j\in S  \tag{LP.3} \label{ILP-constr3}
\end{alignat}

We have introduced the variable $s_i$ to indicate whether $f_i$ is chosen for the cover $A\subseteq S$ ($s_i = 1$ means that $f_i$ is chosen, otherwise $s_i = 0$). We have also introduced the auxiliary variable $t_{ij}$ to indicate whether $f_i$ and $f_j$ are \emph{both} chosen. Hence, constraint \ref{ILP-constr1} ensures that each element $c_k\in U$ is covered by at least one pair in $S$. 
\ref{ILP-constr2} ensures that a pair of elements in $S$ cannot cover any $c_k\in U$ unless both elements are chosen.

To convert the integer program to an \textsc{Ising} instance, we first convert the constraints into expressions of logical operations. \ref{ILP-constr1} can be rewritten as
\begin{equation}\label{eq:cons1}
\bigvee_{c_k\in Q_{ij}}t_{ij}^{(k)}=1,\quad\forall c_k\in U
\end{equation}

\ref{ILP-constr2} can be translated to a truth table for the binary operation involving $t_{ij}$ and $s_i(s_j)$ where only the entry $\{s_i=0,t_{ij}=1\}(s_j=0,t_{ij}=1)$ evaluates to 0 and the other three entries evaluate to 1. Using the following Hamiltonians we could translate the logic operations $\vee$, $\wedge$ and $\le$ into the ground states of Ising model, see \cite{biamonte2008nonperturbative} for more details.

\begin{equation}\label{eq:basicham}
\begin{array}{ccl}
H_{\vee}(s_1,s_2,s_*) & = & \displaystyle \frac{1}{4}(3\ii-\sigma_1^z-\sigma_2^z+2\sigma_*^z+\sigma_1^z\sigma_2^z-2\sigma_1^z\sigma_*^z-2\sigma_2^z\sigma_*^z) \\[0.1in]
H_{\wedge}(s_1,s_2,s_*) & = & \displaystyle \frac{1}{4}(4\ii+\sigma_1^z+\sigma_2^z-2\sigma_*^z+2\sigma_1^z\sigma_2^z-3\sigma_1^z\sigma_*^z-3\sigma_2^z\sigma_*^z) \\[0.1in]
H_{\le}(s_1,s_2) & = & \displaystyle \frac{1}{4}(\ii-\sigma_1^z+\sigma_2^z-\sigma_1^z\sigma_2^z).
\end{array}
\end{equation}
Note that $H_{\le}(s_1,s_2)$ is essentially $|10\rangle\langle 10|_{s_1s_2}$. In other words we are penalizing the only 2-bit string $s_1s_2$ that violates the constraint $s_1\le s_2$.
The ground state subspace of $H_\vee$ is spanned by $\{|s_1s_2s_*\rangle|s_1\vee s_2=s_*,s_1,s_2,s_*\in\{0,1\}\}$. Similarly, the ground state subspace of $H_\wedge$ is spanned by $\{|s_1s_2s_*\rangle|s_1\wedge s_2=s_*\}$ and that of $H_\le$ spanned by $\{|s_1s_2\rangle|s_1\le s_2\}$. 

By linearly combining the above constraint Hamiltonians, we can enforce multiple constraints to hold at the same time. For example, the statement $s_1\vee s_2\wedge s_3=1$ can be decomposed as simultaneously ensuring $s_1\vee s_2=y$, $y\wedge s_3=z$, and $z=1$. In other words we have used auxiliary variables $y$ and $z$ to transform the constraint $s_1\vee s_2\wedge s_3=1$, which involves a clause $s_1\vee s_2\wedge s_3$ of three variables, to a set of constraints involving only clauses of two variables. Then, the Ising Hamiltonian $H=H_\vee(s_1,s_2,y)+H_\wedge(y,s_3,z)+|0\rangle\langle 0|_z$ has its ground state spanned by states $|s_1s_2s_3yz\rangle$ with $s_1$, $s_2$, and $s_3$ satisfying $s_1\vee s_2\wedge s_3=1$. The third term in $H$ ensures that $z=1$ by penalizing states with $|z\rangle=|1\rangle$.

\begin{figure}
	\begin{center}
		\hspace*{-0.5cm}
		\includegraphics[scale=0.82]{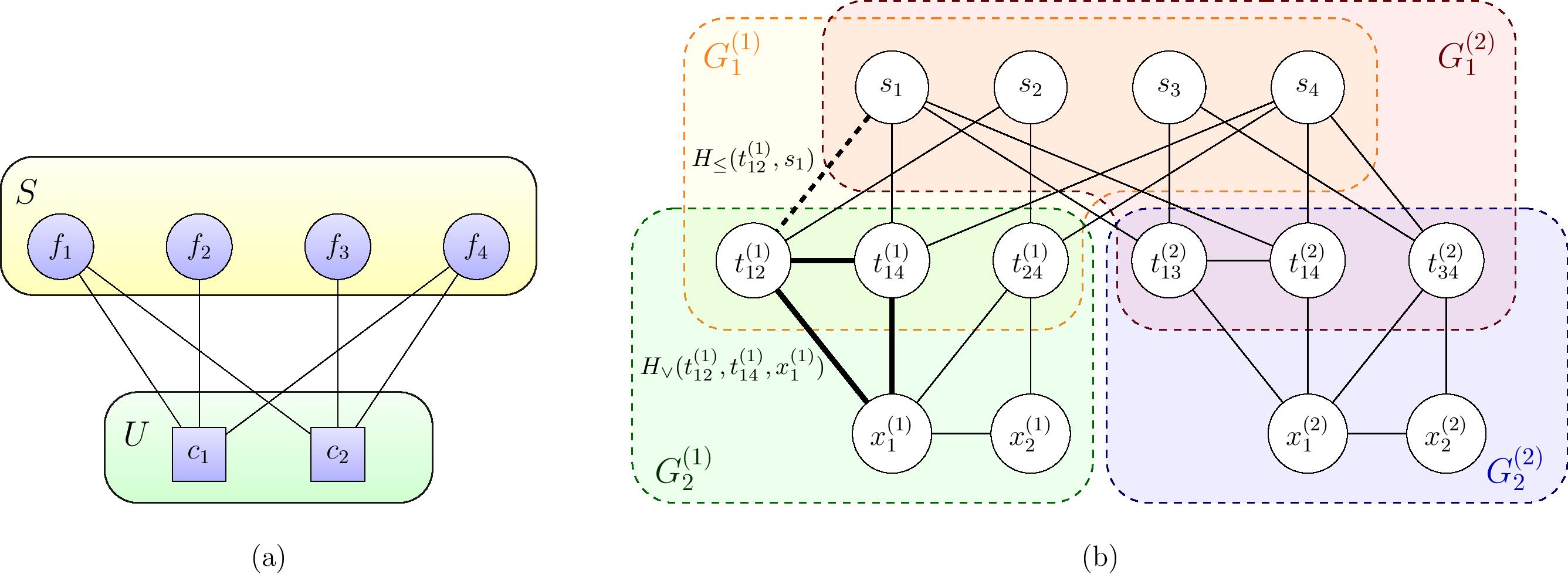}
	\end{center}
	\caption{Example of converting an SCP instance to Ising Hamiltonian. (a) The SCP instance. Here $S=\{f_1,f_2,f_3,f_4\}$ and $U=\{c_1,c_2\}$. The solution is the set $A=\{f_1,f_4\}$. The circles represent the covering set elements $S$ and the squares are the ground elements $U$. (b) The interaction of Ising instance $H_\text{SCP}$ converted from the SCP instance in (a). Every node corresponds to a qubit. The $s_i$'s are the output bits that correspond to the covering set elements $S$. The others are auxiliary variables. Every edge represents an interaction term between the corresponding spins. Here we do not show the 1-local terms in our construction of $H_\text{SCP}$ (for example the terms in $H_\text{targ}$ for enforcing the minimization of the target function). The bold dashed black line exemplifies the edges between the $t_{ij}^{(k)}$ nodes and the $s_i$ nodes, which come from the constraints $t_{ij}^{(k)}\le s_i$ and $t_{ij}^{(k)}\le s_j$ for each pair $\{f_i,f_j\}$ that covers $c_k$. Each of the inequality constraints is enforced by a $H_\le$ term in \eqref{eq:basicham}. The bold triangle exemplifies the $H_\vee$ constraints in \eqref{eq:basicham} that are used to enforce the logical relationship between the $t_{ij}^{(k)}$ variables and the auxiliary variables as shown in \eqref{eq:tij_xi}. The areas marked by $G_1^{(1)}$, $G_1^{(2)}$ etc outline the structure of the Ising Hamiltonian that is relevant in the discussion of hardware embedding.}
	\label{fig:example_SCPP}
\end{figure}

Therefore, we can translate \eqref{eq:cons1} to an Ising Hamiltonian. For a fixed $k$, the constraint \eqref{eq:cons1} takes the form of $t_1^{(k)}\vee t_2^{(k)}\vee\cdots\vee t_{N_k}^{(k)}=1$ where each $t_j^{(k)}\in\{0,1\}$ and $N_k\le\frac{1}{2}m(m-1)=O(m^2)$. Similarly to the example above, we introduce $N_k-1$ auxiliary variables $x_1^{(k)}$, $x_2^{(k)}$, $\cdots$, $x_{N_k-1}^{(k)}$ such that 
\begin{equation}\label{eq:tij_xi}
x_j^{(k)}=\left\{
\begin{array}{cl}
t_1^{(k)}\vee t_2^{(k)} & j=1 \\[0.1in]
x_{j-1}^{(k)}\vee t_{j+1}^{(k)} & j=2,\cdots,N_k-1 \\
\end{array}
\right.
\end{equation}
Thus, $x_{N_k-1}^{(k)}=t_1^{(k)}\vee t_2^{(k)}\vee\cdots\vee t_{N_k}^{(k)}$. In order to ensure the first constrain holds, it is needed to ensure that $x_{N_k-1}^{(k)}=1$. Then we could write down the corresponding Ising Hamiltonian for the constraint as
\begin{equation}
H_k=H_\vee(t_1^{(k)},t_2^{(k)},x_1^{(k)})+\sum_{j=2}^{N_k-1}H_\vee(x_{j-1}^{(k)},t_{j+1}^{(k)},x_j^{(k)})+|0\rangle\langle{0}|_{x_{N_k-1}^{(k)}}.
\end{equation}
The last term is meant to make sure that $x_{N_k-1}^{(k)}=1$ in the ground state of $H_k$. Therefore the Hamiltonian whose ground state subspace is spanned by all states that obey both of the constraints in the integer program \eqref{eq:cons1} can be written as
\begin{equation}
\begin{array}{rcl}
H_\text{cons} & = & \displaystyle \sum_{c_k\in U}H_k + H_\le \\[0.15in]
H_\le & = & \displaystyle \sum_{i,j:f_i,f_j\in S}\left(H_\le(t_{ij},s_i)+H_\le(t_{ij},s_j)\right).
\end{array}
\end{equation}
The target function $\sum_{f_i\in S}s_i$ which we seek to minimize can be directly mapped to an Ising Hamiltonian $H_\text{targ}=\sum_{f_i\in S}|1\rangle\langle 1|_{s_i}=\sum_{f_i\in S}{\frac{1}{2}(1-\sigma_{s_i})}$. This is because we would like to essentially minimize the number of 1's in the set of $s_i$ values and penalize choices with more 1's. Therefore the final Hamiltonian whose ground state contains the solution to the original \textsc{SCP} instance becomes
\begin{equation}\label{eq:hSCPP}
H_\textsc{SCP}=\alpha H_\text{targ}+H_\text{cons}
\end{equation}
for some weight factor $\alpha$. 

We now estimate the overhead for the mapping. $H_\text{targ}$ acts on $|S|=m$ qubits. In $H_\text{cons}$, $H_\le$ acts on $O(m^2)$ qubits, since there are $O(m^2)$ variables $t_{ij}$. Each $H_k$ in $H_\text{cons}$ requires $N_k=O(m^2)$ qubits. There are in total $|U|=n$ of the $H_k$ terms, which gives $O(nm^2)$ qubits in total. $\quad\square$

\subsubsection*{Example}

Consider the SCP instance shown in Figure \ref{fig:example_SCPP}a. With the mapping presented in Theorem \ref{thm:scp_ising}, we arrive at an Ising instance \textsc{Ising}$({\bf h},{\bf J})$ where $\alpha = {1/4}$ in \eqref{eq:hSCPP} and ${\bf h}$, ${\bf J}$ are presented in Supplementary Material. The ground state subspace of the Hamiltonian in \eqref{eq:ising} with $h_i$ and $J_{ij}$ coefficients defined above, restricted to the $s_i$ elements is spanned by $\{|\psi\rangle=|s_1s_2\cdots x_2^{(2)}\rangle\text{ such that }|s_1s_2s_3s_4\rangle = |1001\rangle\}$. This corresponds to $A=\{f_1,f_4\}$, the solution to the SCP instance. Figure \ref{fig:example_SCPP}b illustrates the interaction graph of the spins in the Ising Hamiltonian that corresponds to the SCP instance.

\subsection{Numerical simulation of quantum annealing}

In order to test the time complexity of using quantum annealing to solve SCP instances via the construction in Theorem \ref{thm:scp_ising}, we generate random instances of SCP that lead to Ising Hamiltonian $H_{SCP}$ of $M=3,4,\cdots,19$ spins. In Definition \ref{def:SCPP} we use a bipartite graph between the ground state $U$ of size $n$ and the cover set $S$ of size $m$ to describe an SCP instance. For fixed $n$ and $m$, there are in total $2^{mn}$ such possible bipartite graphs (if we consider each bipartite graph as a subgraph of $K_{m,n}$ and count the cardinality of the power set of the edges of $K_{m,n}$). Therefore to generate random bipartite graphs we only need to flip $mn$ fair coins to uniformly choose from all possibile bipartite graphs between $U$ and $S$. However, we would like to exclude the bipartite graphs where some element of $S$ is not connected to any element in $U$. These ``dummy nodes" are not pertinent to the computational problem at hand and should be removed from consideration before converting the SCP instance to an Ising Hamiltonian $H_\text{SCP}$. We thus use a scheme for generating random instances of SCP \emph{without} dummy nodes as described in Algorithm \ref{alg:dummyfree}. Under the constraint that no dummy element in $S$ is allowed, there are in total $(2^n-1)^m$ possible bipartite graphs. In Supplementary Material we rigorously show that Algorithm \ref{alg:dummyfree} indeed samples uniformly among the $(2^n-1)^m$ possible ``dummy-free" bipartite graphs.

\begin{algorithm}
	\caption{\textbf{Algorithm for generating a random \textsc{SCP}$(G,U,S)$ without dummy elements in the cover set}}
	\label{alg:dummyfree}
	{\bf Input:} The ground set $U$ and the cover set $S$ \newline\newline
		\noindent{\bf Procedure:}
	\begin{algorithmic}[1] 
		
		\State Initialize the output graph $G\leftarrow\emptyset$;
		\ForAll {$s\in S$} \label{step:allins}
		\ForAll {$u\in U$} \label{step:allinu}
		\State With probability 1/2, add edge $(s,u)$ to $G$;
		\EndFor \label{step:allinuend}
		\If {$s$ is still unattached to any element in $U$}
		\State Repeat steps \ref{step:allinu} through \ref{step:allinuend} until $s$ is attached to some element in $U$. \label{step:repeat}
		\EndIf
		\EndFor
		\State {\bf return} $G$.
	\end{algorithmic}
\end{algorithm}

\begin{figure}
\centering
\includegraphics[scale=0.7]{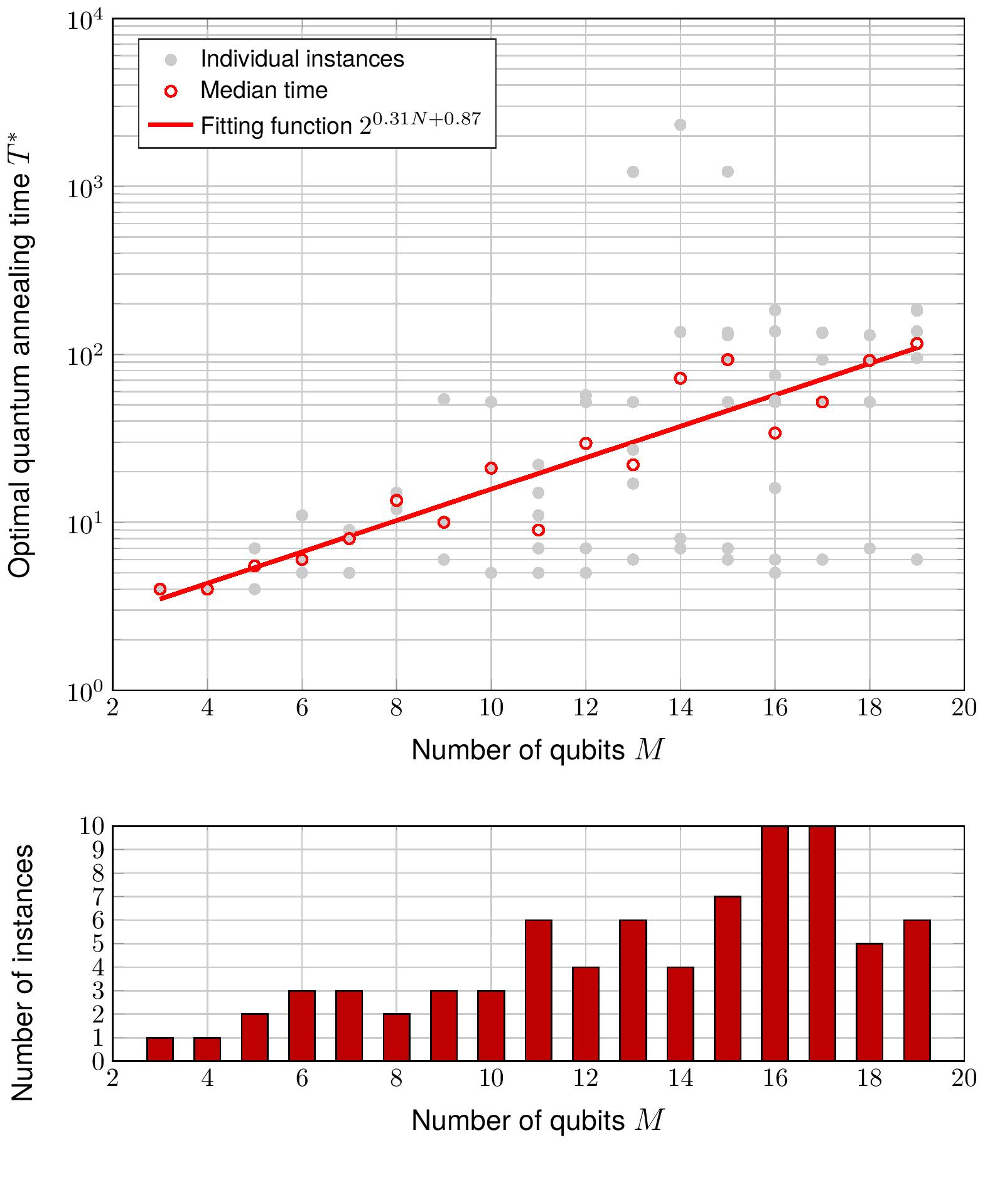}
\caption{Plot of the optimal quantum annealing time $T^*$ versus the number of spins involved in the construction of $H_\text{SCP}$. Here we fit the logarithm of median $T^*$ with a straight line. The size $M$ of our Ising systems ranges from 3 to 19. From the fitting function we observe that the annealing time scales as roughly $O(2^{0.31M})$. We also provide on the bottom plot the number of instances for each $M$.}
\label{fig:runtime_data}
\end{figure}

For each randomly generated instance from Algorithm \ref{alg:dummyfree} we construct an Ising Hamiltonian $H_{SCP}$ according to Theorem \ref{thm:scp_ising}. We then perform a numerical simulation of the time dependent Schr\"{o}dinger equation \eqref{eq:Schr} from time $t=0$ to $t=T$ with time step $\Delta t=1$ and the time dependent Hamiltonian defined as
\begin{equation}
\begin{array}{rcl}
H(s(t)) & = & \displaystyle \left(1-\frac{t}{T}\right)H_B+\frac{t}{T}H_{SCP} \\[0.1in]
H_B & = & \displaystyle \sum_{i=1}^M\sigma_i^x \\[0.1in]
\end{array}
\end{equation}
where $H_\text{SCP}$ is defined in equation \eqref{eq:hSCPP}. Here because of the construction of $H_\text{SCP}$, our total Hamiltonian $H(s(t))$ acts not only on the spins ${\bf s}\in\{0,1\}^m$ indicating our choice of elements in the cover set $S$, but also auxiliary variables $t_{ij}^{(k)}$ and $x_i^{(k)}$, for which we use ${\bf t}$ and ${\bf x}$ to denote their respective collections. Our initial state is the ground state of $H_B$, namely 
\begin{equation}
|\psi(0)\rangle=\frac{1}{\sqrt{2^M}}\sum_{{\bf s},{\bf t},{\bf x}\in\{0,1\}^M}|{\bf s},{\bf t},{\bf x}\rangle.
\end{equation} 
To obtain the final state $|\psi(T)\rangle$ where $T$ is some positive integer, we use the {\tt ode45} subroutine of MATLAB under default settings to numerically integrate Schr\"{o}dinger equation to obtain $|\psi(1)\rangle$ from $|\psi(0)\rangle$, and then use $|\psi(1)\rangle$ as an initial state to obtain $|\psi(2)\rangle$ in the same fashion, and so on. We define the success probability $p$ as a function of the total annealing time $T$ as $p(T)=\|\Pi|\psi(T)\rangle\|_2$ where $\Pi$ is a projector onto the subspace spanned by states with $\bf s$ being a solution of the original SCP instance. Using binary search we determine the minimum time $T^*$ to achieve $p(T^*)\ge 0.25$ for each instance of SCP. Figure \ref{fig:runtime_data} shows the distribution of $T^*$ for SCP instances that lead to Ising Hamiltonians $H_\text{SCP}$ of the same sizes, as well as how the median annealing time scales as a function of number of spins $M$. Results show that for instances with $M$ up to 19, the median annealing time scales roughly as $O(2^{0.31M})$.

\subsection{Numerical experiment with Simulated Annealing}\label{sec:sacompare}

Simulated annealing, first introduced three decades ago\cite{KGV83}, has been widely used as a heuristic for handling hard combinatorial optimization problems. It is especially of interest as a benchmark for quantum annealing\cite{FGG02,SMTC02,HRIT14} because of similarities between the two algorithms. While quantum annealing employs quantum tunneling to escape from local minima, simulated annealing relies on thermal excitation to avoid being trapped in local minima. The general procedure we adopt for simulated annealing to approach the ground state of an Ising spin glass can be summarized as the following\cite{IZRT}: 
\begin{enumerate}
\item Repeat $R$ times the following:
\begin{enumerate}
\item Initialize ${\bf s}\leftarrow{\bf s}_0$ randomly;
\item Perform $S$ times the following: (let $i=0,1,\cdots,S-1$ index the steps)
\begin{enumerate}
\item Set the temperature $T_i\leftarrow \tau(i)$;
\item Perform a \emph{sweep} on ${\bf s}_i$ to obtain ${\bf s}'$; (a sweep is a sequence of steps each of which randomly selects a spin and flips its state, so that on average each spin is flipped once during a sweep)
\item With probability $\text{exp}(\frac{E({\bf s}')-E({\bf s})}{T_i})$, let ${\bf s}_{i+1}={\bf s}'$. Otherwise let ${\bf s}_{i+1}\leftarrow{\bf s}_i$.
\end{enumerate}
\end{enumerate}
\item Return ${\bf s}_S$ as the answer.
\end{enumerate}

For the purpose of comparison we also used simulated annealing to solve the same set of instances generated by Algorithm \ref{alg:dummyfree} for testing quantum annealing. The program implementation that we use is built by Isakov \emph{et al}\cite{IZRT}, which is a highly optimized implementation of simulated annealing with care taken to exploit the structures of the interaction graph, such as being bipartite and of bounded degree. Here we use the program's most basic realization of single-spin code for general interactions with magnetic field on an interaction graph of any degree.

\begin{figure}[!htb]
\centering
\hspace*{-1.4cm}
\includegraphics[scale=0.7]{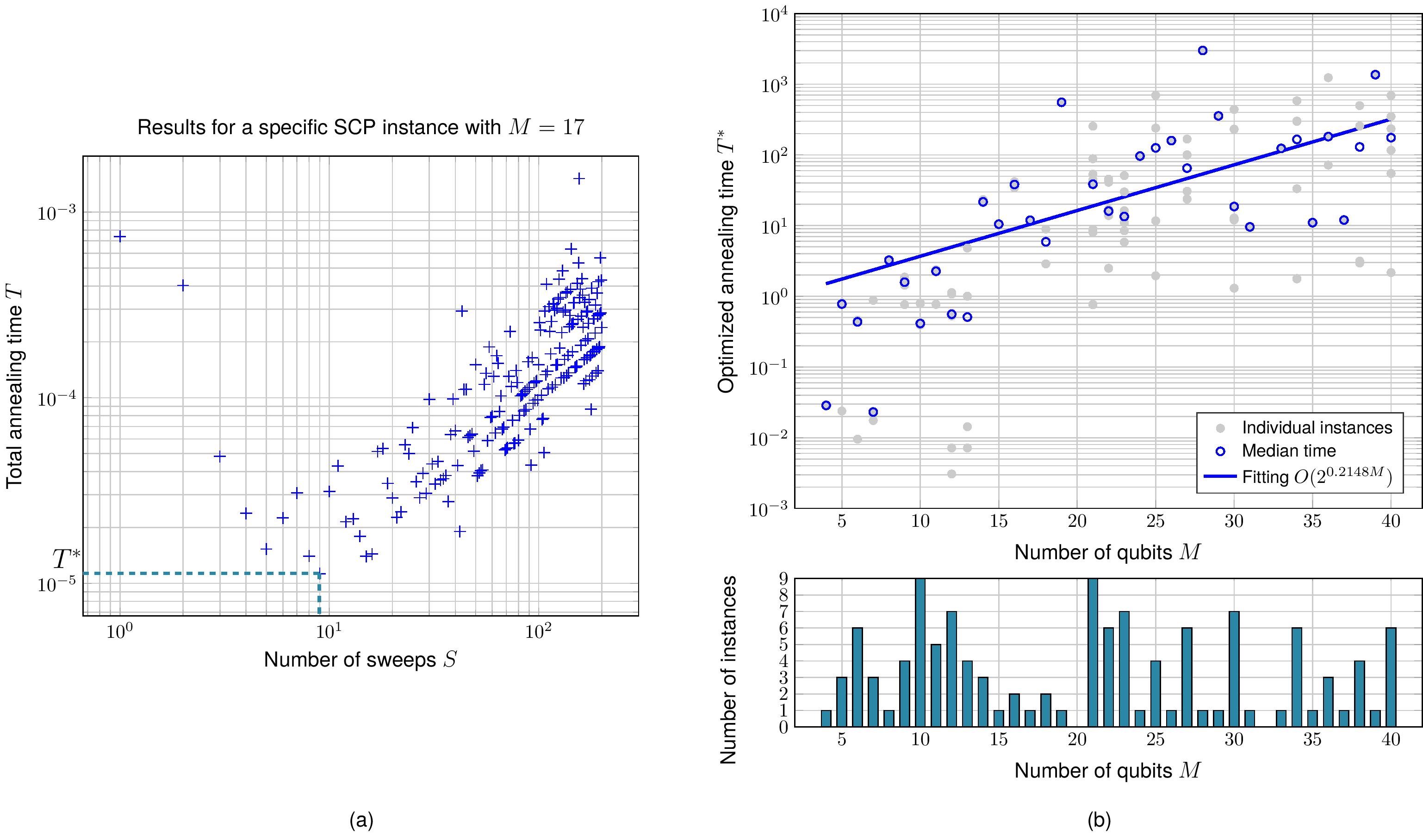}
\caption{(a) Plot of annealing time $T$ versus number of sweeps $S$ using the simulated annealing implementation\cite{IZRT} on an Ising Hamiltonians of 17 spins constructed from an SCP instance. We use the default settings for all parameters other than $S$ and $R$. Also we mark the optimal runtime $T_*$. (b) Plot of optimized annealing time $T^*$ versus the number of spins involved in the Ising Hamiltonian HSCP corresponding to randomly generated SCP instances according to Algorithm 1. We also provide on the bottom plot the number of instances for each $M$.}
\label{fig:time_sweep}
\end{figure}

As mentioned by Isakov \emph{et al.}, to improve the solution returned by simulated annealing, one could increase either the number of sweeps $S$ or number of repetitions $R$ in the implementation, or both of them. However, note that the total annealing time is proportional to the product $S\cdot R$ and there is a trade-off between $S$ and $R$. For a fixed number of sweeps $S$ let the success probability (\emph{i.e. }the fraction of ${\bf s}_i$ that is satisfactory) be $w(S)$. In order to achieve a constant success probability $p$ (say 25\%, which is what we use here), we need at least $R=\lceil\log(1-p)/\log(1-w(S))\rceil$ repetitions. Hence the total time of simulated annealing can be written as
\begin{equation}
T(S)=\left\lceil\frac{\log(1-p)}{\log(1-w(S))}\right\rceil\cdot S.
\end{equation}
In general $w(S)$ increases as $S$ increases, leading to a decrease in $R$. We numerically investigate this with an Ising system of $N=17$ spins generated from an SCP instance via the construction in Theorem \ref{thm:scp_ising}. We plot the annealing time $T$ versus $S$ in Figure \ref{fig:time_sweep}a. For each SCP instance with the number of spin $M$ we compute the optimal $S^*$ such that $T^*=T(S^*)$ is the optimized runtime (Figure \ref{fig:time_sweep}a). We further explore how the optimal runtime $T^*$ scales as a function of the number of spins $M$. As shown in Figure \ref{fig:time_sweep}b, a linear fit on a semilog plot shows that roughly $T^*=O(2^{0.21M})$. 

The units of time used for both Figure \ref{fig:time_sweep}a and Figure \ref{fig:time_sweep}b are arbitrary and thus do not support a point-to-point comparison. But the scaling difference seems apparent. For quantum annealing we restrict to systems of at most 19 spins due to computational limitations faced in representing the full Ising Hamiltonian when numerically integrating the time-dependent Schr\"{o}dinger equation \eqref{eq:Schr}. 

Although there is no quantum speedup observed in terms of median runtime over all randomly generated instances of the same size, we notice that for a fixed number of spins $M$ the performances of both quantum annealing and simulated annealing are sensitive to the specific instance of Ising Hamiltonian $H_\text{SCP}$ than simulated annealing. This can be seen by considering at the same time the quantum annealing results in Figure \ref{fig:runtime_data} and the test results for simulated annealing shown in Figure \ref{fig:time_sweep}b. One could then speculate that perhaps by focusing on a specific subset of SCP instances could yield a quantum advantage. 

\section{Embedding on quantum hardware} \label{subsec:embed_chimera}

In this section we deal with the physical realization of quantum annealing for solving SCP instances using D-Wave type hardwares. There are mainly two aspects\cite{choi08,choi11} of this effort: 1) The \emph{embedding problem}\cite{choi11}, namely embedding the interaction graph of the Ising Hamiltonian construction $H_\text{SCP}$ as a graph minor of a Chimera graph (refer to Section \ref{subsec:graph_minor} for definitions of the graph terminologies). 2) The \emph{parameter setting problem}\cite{choi08}, namely assigning the strengths of the couplings and local magnetic fields for embedded graph on the hardware, in a way that minimizes the energy scaling (or control precision) required for implementing the embedding. Here we focus on the former issue.

We start with an observation on the structures of $H_\text{SCP}$. For any instance \textsc{SCP}$(G,U,S)$ according to Definition \ref{def:SCPP}, the interaction graph $I_\text{SCP$(G,U,S)$}$ of the corresponding Ising Hamiltonian $H_\text{SCP}$ can be regarded as a union of $n$ subgraphs, namely $I_\text{SCP$(G,U,S)$}=G^{(1)}\cup G^{(2)}\cup\cdots\cup G^{(n)}$. Each subgraph $G^{(i)}$ is associated with an element of the ground set $c_i\in U$ as in Figure \ref{fig:example_SCPP}a. Each $G^{(i)}$ could be further partitioned into two parts, $G_1^{(i)}$ and $G_2^{(i)}$. For any $k$, $G_1^{(k)}$ is a bipartite graph between $\{s_i\}_{i=1}^m$ and $\{t_{ij}^{(k)}|\text{$f_i$, $f_j\in S$ cover $c_k\in U$}\}$. $G_2^{(k)}$ essentially describes the interaction between the auxiliary variables $t_{ij}^{(k)}$ and $x_i^{(k)}$ as described in equation \eqref{eq:tij_xi}. In Figure \ref{fig:example_SCPP}b we illustrate such partition using the example from Figure \ref{fig:example_SCPP}b. Our goal is then to show constructively that $I_\text{SCP$(G,U,S)$}\le_m F(f_1,f_2,c)$ for some $f_1$, $f_2$ that depend on $m$, $n$ and $c=4$, which describes the Chimera graph realized by D-Wave hardware (Figure \ref{fig:chimera_notation}a).

It is known\cite{KSH12} that one could embed a complete graph on $cm+1$ nodes onto Chimera graph $F(m,m,c)$. Since any $n$-node graph is a subgraph of the $n$-node complete graph, in principle any $n$-node graph can be embedded onto Chimera graphs of size $O(n^2)$ using the complete graph embedding. A downside of this approach is that it may fail to embed many graphs that are in fact embeddable\cite{KSH12}. Also, using embeddings based on complete graph embeddings will likely lose the intuition on the structure of the original graph. For graphs with specific structures, such as bipartite graphs one may be able to find an embedding that is also in some sense structured. We show in the following Lemma an embedding for any complete bipartite graph $K_{p,q}$ onto a Chimera graph. The ability to do so enables us to embed any bipartite graph onto a Chimera graph.

\begin{lemma}\label{lem:bipart_embed}
For any positive integers $p$, $q$ and $c$, $K_{p,q}\le_mF(\lceil q/c\rceil,\lceil p/c\rceil, c)$.
\end{lemma}
\begin{proof}
By the definition of graph minor embedding in Section \ref{subsec:graph_minor}, it suffices to construct a mapping $\phi_{p,q}:K_{p,q}\mapsto F(\lceil q/c\rceil,\lceil p/c\rceil, c)$ where each $v$ in $F_{p,q}$ is mapped to a tree $T_v$ in $F(\lceil q/c\rceil,\lceil p/c\rceil, c)$ and each edge $e=(u,v)$ in $K_{p,q}$ is mapped to an edge $(i_u,i_v)$ with $i_u\in T_u$ and $i_v\in T_v$. 

Let $i=1,2,\cdots,p$ label the nodes on one side of $K_{p,q}$ and $j'=1,2,\cdots,q$ label the nodes in the other. Using the labelling scheme on the nodes of Chimera graphs introduced in Section \ref{subsec:chimera} and Figure \ref{fig:chimera_notation}b, we define our mapping $\phi_{p,q}$ as
\begin{equation}\label{eq:phi_map}
\begin{array}{ccl}
\phi_{p,q}(i) & = & \{v_{i\text{ mod } c}^{(t,\lceil i/c\rceil)}|t=1,\cdots,\lceil q/c\rceil\} \\[0.1in]
\phi_{p,q}(j') & = & \{v_{c+(j'\text{ mod } c)}^{(\lceil j'/c\rceil,t)}|t=1,\cdots,\lceil p/c\rceil\}. \\[0.1in]
\phi_{p,q}(i,j') & = & \left(v_{i\text{ mod }c}^{(\lceil i/c\rceil,\lceil j'/c\rceil)}, v_{c+(j'\text{ mod }c)}^{(\lceil i/c\rceil,\lceil j'/c\rceil)}\right)
\end{array}
\end{equation}
where $\phi_{p,q}(u,v)$ maps an edge $(u,v)$ in $K_{p,q}$ to the Chimera graph. If we choose the edges in the Chimera graph properly, it could be checked that $\phi_{p,q}(K_{p,q})$ is a subgraph of $F(\lceil q/c\rceil,\lceil p/c\rceil, c)$.
\end{proof}

\begin{figure}
\centering
\includegraphics[scale=0.8]{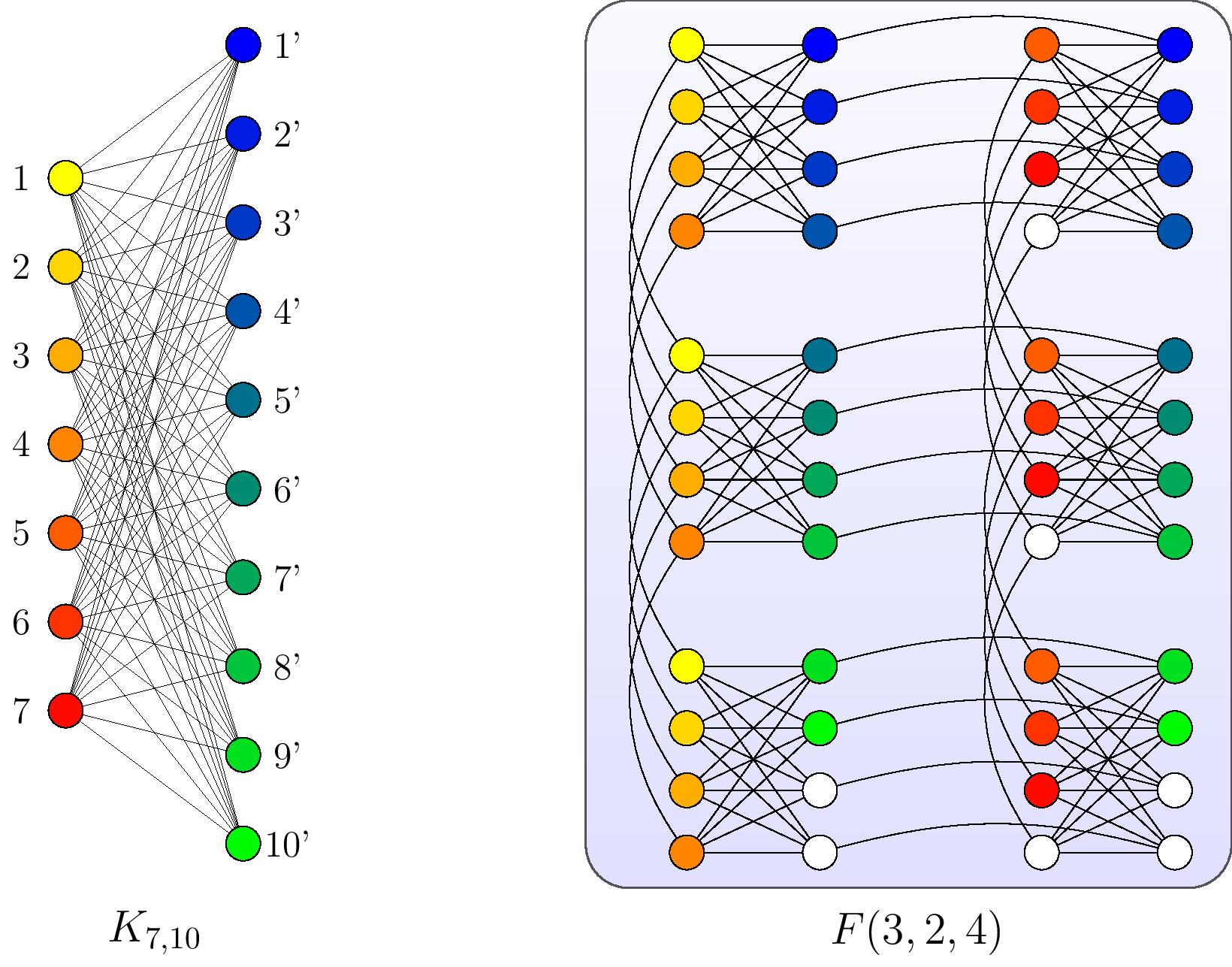}
\caption{An example showing the embedding scheme outlined in Lemma \ref{lem:bipart_embed}. The nodes and the trees mapped from the nodes are marked with the same colors.}
\label{fig:K510}
\end{figure}

In Figure \ref{fig:K510} we show an example of embedding $K_{7,10}$ into $F(3,2,4)$. A natural corollary of Lemma \ref{lem:bipart_embed} is that any bipartite graph between $p$ and $q$ nodes can be minor embedded in $F(\lceil q/c\rceil, \lceil p/c\rceil, c)$. We are then prepared to handle embedding the $G_1^{(i)}$ parts of the interaction graphs of $H_\text{SCP}$, which are but bipartite graphs (see Figure \ref{fig:example_SCPP}b for example).

\begin{figure}
\centering
\includegraphics[scale=0.9]{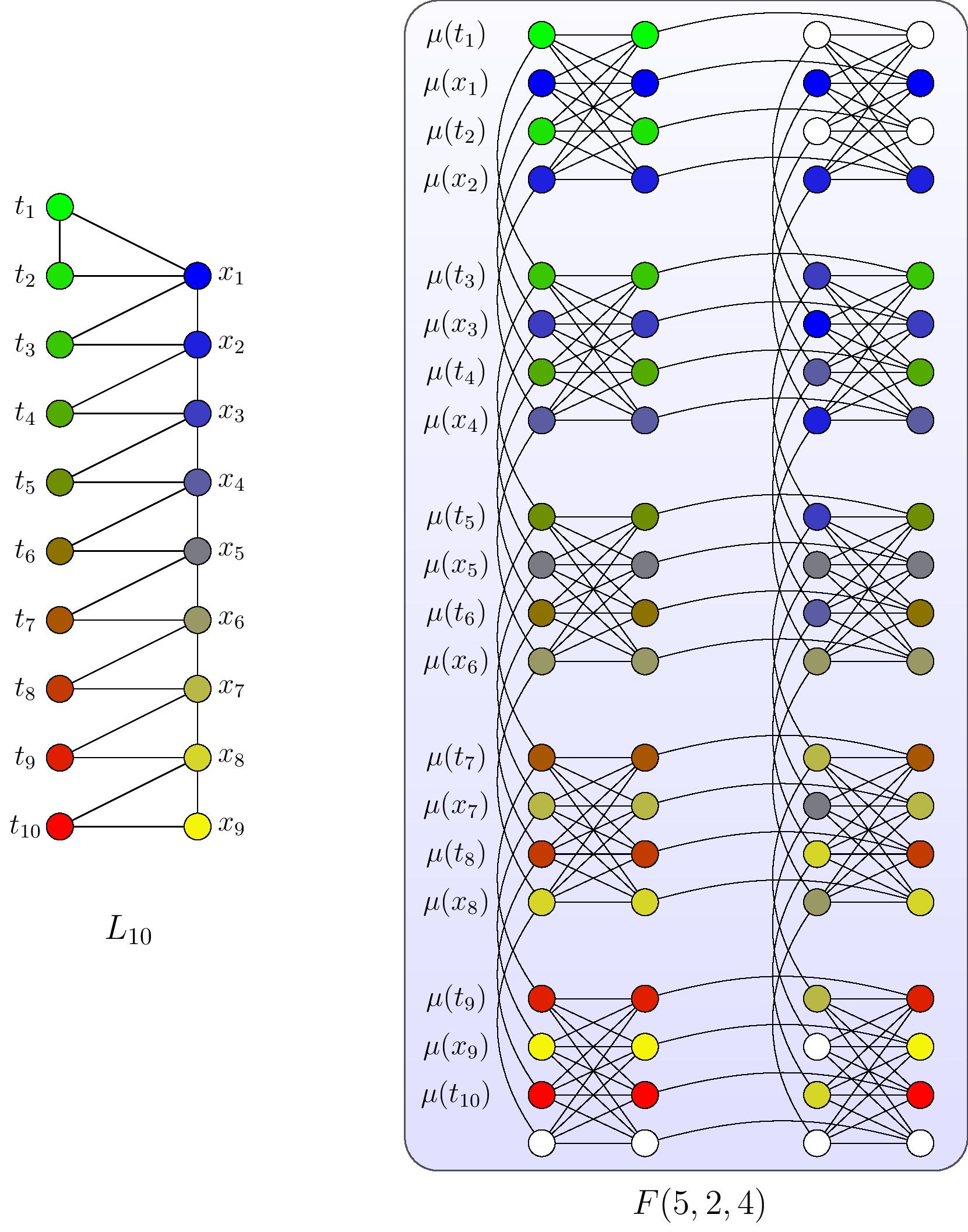}
\caption{An example of embedding $L_{10}$ onto $F(5,2,4)$. Each color in the left diagram represents a node $u$ in $L_{10}$ and the nodes of the same color in the right diagram shows $\mu_{10}(u)$.}
\label{fig:Hn_example}
\end{figure}

We then proceed to treat the $G_2^{(i)}$ parts of the interaction graph. The connectivity of $G_2^{(k)}$ is completely specified by \eqref{eq:tij_xi}. To describe such connectivity we define a family of graph $L_n(V_n,E_n)$ as $V_n=T_n\cup X_{n-1}$ where $T_n=\{t_1,t_2,\cdots,t_n\}$ and $X_{n-1}=\{x_1,x_2,\cdots,x_{n-1}\}$ are two disjoint sets of nodes, the former representing the intermediate variables $t_{ij}^{(k)}$ and the latter representing the $x_k$ variables in equation \eqref{eq:tij_xi}. The set of edges takes the form 
\begin{equation}
E_n=\{(t_1,t_2),(t_1, x_1),(t_2,x_1)\}\cup\left(
\bigcup_{i=2}^{n-1}\{(x_{i-1},x_i),(x_{i-1},t_{i+1}),(x_i,t_{i+1})\}
\right).
\end{equation}
In Figure \ref{fig:Hn_example} we show an example of $L_{10}$. For any $k=1,2,\cdots,|U|$, let $r_k$ be the number of pairs $f_i,f_j\in S$ that cover $k$. Then $G_2^{(k)}=L_{r_k}$. Hence in order to show that we could embed any $G_2^{(i)}$ onto a Chimera graph, it suffices to show that we can embed any $L_n$ onto a Chimera graph. We show this in the following Lemma for $c=4$.

\begin{lemma}\label{lemma:Ln}
For any positive integer $n$, $L_n\le_mF(\lceil{2n/c}\rceil,2,c)$ where we restrict to $c=4$.
\end{lemma}
\begin{proof}
Similar to Lemma \ref{lem:bipart_embed}, we construct a mapping $\mu_n:L_n\mapsto F(\lceil{2n/c}\rceil,2,c)$ where we fix $c=4$. Following the notation for nodes in $L_n$ in Figure \ref{fig:Hn_example} and the notation for nodes in $F(p,q,c)$ in Figure \ref{fig:chimera_notation}b, we construct $\mu$ as
\begin{equation}
\begin{array}{ccl}
\mu_n(t_i) & = & \{v_{(2i-1)\text{ mod }c}^{(\lceil\frac{2i-1}{c}\rceil,1)},v_{c+[(2i-1)\text{ mod }c]}^{(\lceil\frac{2i-1}{c}\rceil,1)}\}\cup\xi_t(t_i) \\[0.1in]
\mu_n(x_i) & = & \{v_{(2i)\text{ mod }c}^{(\lceil\frac{2i}{c}\rceil,1)},v_{c+[(2i)\text{ mod }c]}^{(\lceil\frac{2i}{c}\rceil,1)},v_{c+[(2i)\text{ mod }c]}^{(\lceil\frac{2i}{c}\rceil,2)}\}\cup\xi_x(x_i)
\end{array}
\end{equation}
where $\xi_t$ and $\xi_x$ are defined as
\begin{equation}
\xi_t(t_i)=\left\{
\begin{array}{ll}
\emptyset & \text{if $i=1,2$} \\[0.1in]
\{v_{c+[(2i-1)\text{ mod }c]}^{(\lceil\frac{2i-1}{c}\rceil,2)}\} & \text{otherwise}.
\end{array}\right.
\end{equation}
\begin{equation}
\xi_x(x_i)=\left\{
\begin{array}{ll}
\{v_{(2i)\text{ mod }c}^{(\lceil\frac{2i}{c}\rceil,2)},v_{(2i)\text{ mod }c}^{(\lceil\frac{2i}{c}\rceil+1,2)}\} & \text{if $\lceil i/2\rceil\text{ mod }2=1$ and $2i+4<2n-1$}  \\ [0.1in]
\{v_{[(2i)\text{ mod }c] - 1}^{(\lceil\frac{2i}{c}\rceil,2)},v_{[(2i)\text{ mod }c] - 1}^{(\lceil\frac{2i}{c}\rceil+1,2)}\} & \text{if $\lceil i/2\rceil\text{ mod }2=0$ and $2i+4<2n-1$} \\ [0.1in]
\emptyset & \text{otherwise}.
\end{array}\right.
\end{equation}
With the vertex mapping $\mu_n$, a mapping of edges in $L_{n}$ onto the Chimera graph $F(\lceil 2n/c\rceil,2,c)$ is easy to find.
\end{proof}

In Figure \ref{fig:Hn_example} we show an example of embedding $L_{10}$ onto $F(5,2,4)$. We could then proceed to embed the interaction graph $I_\text{SCP$(G,U,S)$}$, such as the one shown in Figure \ref{fig:example_SCPP}b, in a Chimera graph. Specifically, we state the following theorem.

\begin{figure}
\centering
\includegraphics[scale=0.9]{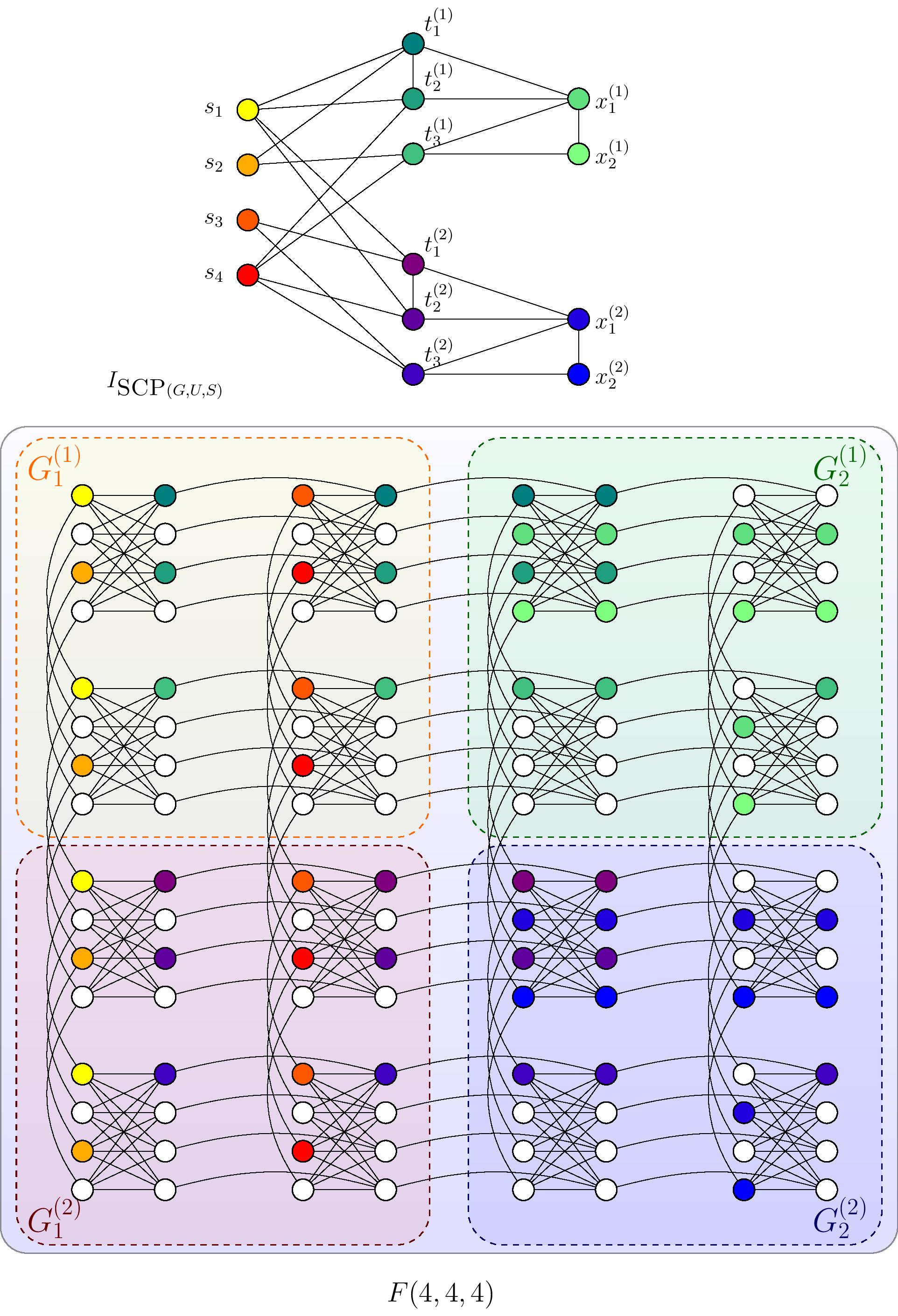}
\caption{Embedding the interaction graph of the example physical system in Figure \ref{fig:example_SCPP}b onto $F(4,4,4)$. Note that the structure of Figure \ref{fig:example_SCPP} is preserved on the Chimera graph.}
\label{fig:Iscp}
\end{figure}

\begin{theorem}\label{thm:embedding}
For any instance $\textsc{SCP}(G,U,S)$ with $|U|=n$ and $|S|=m$, $I_{\textsc{SCP}(G,U,S)}\le_mF(f_1,f_2,c)$ where $f_1=O(nm^2)$, $f_2=O(m)$ and $c=4$ is a constant.
\end{theorem}
\begin{proof}
Our embedding combines ideas from Lemma \ref{lem:bipart_embed} and \ref{lemma:Ln}. We modify the mapping $\phi_{p,q}$ constructed in Lemma \ref{lem:bipart_embed} to produce a new mapping $\theta_{p,q}$ that produces more spacing between the embedded nodes (see for example $G_1^{(1)}$ and $G_1^{(2)}$ in Figure \ref{fig:Iscp}):
\begin{equation}\label{eq:phi_map}
\begin{array}{ccl}
\theta_{p,q}(i) & = & \{v_{(2i-1)\text{ mod } c}^{(t,\lceil (2i-1)/c\rceil)}|t=1,\cdots,\lceil q/c\rceil\} \\[0.1in]
\theta_{p,q}(j') & = & \{v_{c+(2j'-1 \text{ mod } c)}^{(\lceil (2j'-1)/c\rceil,t)}|t=1,\cdots,\lceil p/c\rceil\} \\[0.1in]
\end{array}
\end{equation}
Let $\mu_n^{(r,s)}$ denote a mapping $\mu$ described in Lemma \ref{lemma:Ln} that maps the upper left node (Figure \ref{fig:Hn_example}) $t_1$ to $v_1^{(r,s)}$ instead of $v_1^{(1,1)}$. The rest of the mapping then proceeds from $v_1^{(r,s)}$. In other words, $\mu_n^{(r,s)}$ is the mapping $\mu$ that is shifted by $p-1$ cells to the right and $q-1$ cells below. Trivially $\mu_n^{(1,1)}=\mu$. Similarly we define $\theta_{p,q}^{(r,s)}$ as the shifted embedding under $\theta_{p,q}$ where $\theta_{p,q}^{(r,s)}(1)= v_1^{(r,s)}$. Recall that for any ground set element  $c_k\in U$, $r_k$ is the number of pairs in $S$ that covers $c_k$. We could then specify the embedding from $I_{\textsc{SCP}(G,U,S)}$ onto $F(f_1,f_2,c)$ as
\begin{equation}
\begin{array}{ccl}
\Phi(s_i) & = & \displaystyle \bigcup_{j=1}^n\theta_{m,r_i}^{(1+d_j,1)}(i),\text{  where $i=1,\cdots,m$} \\[0.1in]
\Phi(t_i^{(j)}) & = & \displaystyle \theta_{m,r_i}^{(1+d_j,1)}(t_i)  \cup \mu_{r_j}^{(1+d_j,1+\lceil 2m/c\rceil)}(t_i),\text{  where $i=1,\cdots,r_j$ and $j=1,\cdots,n$}\\[0.1in]
\Phi(x_i^{(j)}) & = & \displaystyle \mu_{r_j}^{(1+d_j,1+\lceil 2m/c\rceil)}(t_i),\text{ where $i=1,\cdots,r_j-1$ and $j=1,\cdots,n$}
\end{array}
\end{equation}
where $d_j=\sum_{k=1}^{j-1}\lceil{2r_k}/{c}\rceil$ is the total number of rows of cells occupied by the embedded graphs for handling the ground elements $c_1$ through $c_{j-1}$. In total $\Phi(I_\text{SCP$(G,U,S)$})$ will occupy $f_1=\sum_{k=1}^n\lceil 2r_k/c\rceil\le n \left(\begin{smallmatrix}m \\ 2\end{smallmatrix}\right)\cdot{2}/{c}=O(nm^2)$ rows and $f_2=\lceil{2m/c}\rceil+2=O(m)$ columns.
\end{proof}

In Figure \ref{fig:Iscp} we show an embedding $I_\text{SCP(G,U,S)}$ of the example instance in Figure \ref{fig:example_SCPP} onto $F(4,4,4)$. Note that our embedding preserves the original structure of the interaction graph as shown in Figure \ref{fig:example_SCPP}b. Furthermore, note that the interaction graph $I_\text{SCP$(G,U,S)$}$ has $M=O(nm^2)$ nodes. Using the complete graph embedding requires $O(M^2)=O(n^2m^4)$ qubits. For the same reason, the construction of Ising Hamiltonian described in equation \eqref{eq:ising_lucas} is likely going to cost $O(nm^4)$ in the worst case of embeding in a Chimera graph since the interaction graph of the Hamiltonian could involve complete graphs of size $O(m^2)$ due to the square term $H_A$. By comparison our embedding costs $f_1f_2\cdot 2c=O(nm^3)$ qubits and preserves the structure of the original instance, which affords slightly more advantage for scalable physical implementations.

\section{Discussion}

Our interest in SCP is largely motivated by its important applications in various areas\cite{BDD+11,LPR04,WY11,gonccalves2012exact}. We have shown a complete pipeline of reductions that converts an arbitrary SCP instance to an interaction graph on a D-Wave type hardware based on Chimera graphs, in a way that preserves the structure of the instance throughout (Figure \ref{fig:example_SCPP}b and \ref{fig:Iscp}) and is more qubit efficient than the usual approach by complete graph embedding. Although no quantum speedup is observed at this stage based on comparison of median annealing times, the large variance of runtimes observed in Figure \ref{fig:runtime_data}a from instance to instance might suggest that specific subsets of instances could provide quantum speedup. Of course, a clearer understanding of the performance of quantum annealing on solving SCP could only be brought forth by both scaling up the numerical simulation of the quantum annealing process to include instances with larger number of spins and actual experimental implementation of the quantum annealing process. Both of them are of interest to us in our future work.

\bibliography{wedet}

\section*{Acknowledgements}
The authors thank Sergei Isakov for helpful discussions on the simulated annealing code, and Howard J. Karloff for the original discussion on the disjoint path facility location problem.

\section{Appendix}
\subsection{Details of the example SCP instance}\label{app:ex_scp}

In the paper we consider an example SCP instance for illustrating our mappings from SCP to Ising and eventually to a Chimera graph. Specifically, the \textsc{Ising}$({\bf h},{\bf J})$ described in Figure \ref{fig:example_SCPP}b has
\[{\bf h}^T = \frac{1}{8}\cdot \bordermatrix{
	~ & s_1 & s_2 & s_3 & s_4 & t_{12}^{(1)} & t_{14}^{(1)} & t_{24}^{(1)} & t_{13}^{(2)} & t_{14}^{(2)} & t_{34}^{(2)} & x_1^{(1)} & x_2^{(1)} & x_1^{(2)} & x_2^{(2)} \cr
	~ & 7 & 3 & 3 & 7 & -6 & -6 & -6 & -6 & -6 & -6 & 2 & 4 & 2 & 4}.\]
Here the labels above each element of ${\bf h}$ indicates which spin the coefficient is associated to. The matrix of interaction coefficients {\bf J} is shown in Figure \ref{fig:J_example}.

\begin{figure}
\begin{equation}\label{eq:J}
\begin{array}{ccl}
{\bf J} & = & \displaystyle\frac{1}{8}\cdot
\bordermatrix{
	~ & s_1 & s_2 & s_3 & s_4 & t_{12}^{(1)} & t_{14}^{(1)} & t_{24}^{(1)} & t_{13}^{(2)} & t_{14}^{(2)} & t_{34}^{(2)} & x_1^{(1)} & x_2^{(1)} & x_1^{(2)} & x_2^{(2)} \\[0.02in]
	s_1 &       &     &    &      & -1  &  -1  &      & -1  &  -1 &       &     &    &      &   \\[0.02in]
	s_2 &       &   &     &     &  -1  &   &   -1  &     &      &    &     &    &     &    \\[0.02in]
	s_3 &      &     &     &     &    &     &     &  -1  &     & -1  &     &      &    &    \\[0.02in]
	s_4 &      &     &     &     &      & -1  &  -1   &    &  -1  &  -1   &    &      &    &    \\[0.02in]
	t_{12}^{(1)} &   -1  &  -1 &       &    &    &   1   &    &     &    &      &  -2  &    &      &    \\[0.02in]
	t_{14}^{(1)} &   -1  &     &      & -1  &   1 &     &      &      &   &     &  -2   &   &       &   \\[0.02in]
	t_{24}^{(1)} &     &   -1  &    &   -1  &     &     &    &     &     &     &   1 &   -2  &     &    \\[0.02in]
	t_{13}^{(2)} &   -1  &    &  -1   &    &      &    &     &     &  1  &     &     &     &  -2  &    \\[0.02in]
	t_{14}^{(2)} &   -1   &    &     &  -1   &     &     &    &   1  &    &      &    &     &  -2  &    \\[0.02in]
	t_{34}^{(2)} &      &     &  -1  &  -1 &      &      &    &     &     &     &    &     &   1  &  -2 \\[0.02in]
	x_1^{(1)} &      &     &      &     & -2 &   -2   &  1  &     &      &    &      & -2  &      &   \\[0.02in]
	x_2^{(1)} &      &    &      &   &     &    &   -2  &      &    &     &  -2  &     &      &   \\[0.02in]
	x_1^{(2)} &      &     &     &     &     &      &     & -2  &  -2  &   1  &      &    &    &   -2 \\[0.02in]
	x_2^{(2)} &      &     &     &    &      &     &    &    &      & -2  &     &    &   -2 &    }
\end{array}
\end{equation}
\caption{The matrix of coupling coefficients in the Ising Hamiltonian instance constructed for the example SCP instance shown in Figure \ref{fig:example_SCPP}. The interpretation of the matrix elements of $\bf J$ follows Definition \ref{def:ising}.}
\label{fig:J_example}
\end{figure}

\subsection{Proof of correctness for Algorithm \ref{alg:dummyfree}}\label{app:nodummy}

Here we show that Algorithm \ref{alg:dummyfree} indeed samples uniformly from all $(2^n-1)^m$ possible ``dummy-free" bipartite graphs for a fixed setting of the ground set $U$ of size $n$ and cover set $S$ of size $m$. Formally we say a bipartite graph $G(U\cup S,E)$ between two sets $U$ and $S$ is \emph{dummy-free} if for any $s\in S$ there exists at least one $u\in U$ such that $(s,u)\in E$. Then we state the following claim.


\begin{claim}
Given any set $U$ of $n$ elements and $S$ of $m$ elements, for any dummy-free bipartite graph $G(V,E)$ between $U$ and $S$, Algorithm \ref{alg:dummyfree} generates $G$ with probability $(2^n-1)^{-m}$. 
\end{claim}
\begin{proof}
Let $\text{Pr}(G)$ be the probability that Algorithm \ref{alg:dummyfree} generates $G$. Recall that if at a particular $s\in S$ during the looping on line \ref{step:allins}, when Algorithm \ref{alg:dummyfree} scanned through all $u\in U$ but did not end up selecting any element in $U$, the algorithm enters line \ref{step:repeat} to repeat the process from scratch for $s$. Then depending on how many times the algorithm entered line \ref{step:repeat} during the process of generating $G$, we could express $\text{Pr}(G)$ as
\begin{equation}\label{eq:prG}
\text{Pr}(G)=\sum_{k=0}^\infty \text{Pr}(G|\text{Algorithm \ref{alg:dummyfree} entered line \ref{step:repeat} in total $k$ times})
\end{equation}
If the algorithm never entered line \ref{step:repeat} and generated $G$, then the probability of generating $G$ is essentially the probability of $mn$ coin flips, namely $2^{-mn}$. If the algorithm entered line \ref{step:repeat} once, then the probability Pr$(G)=2^{-mn}\cdot m2^{-n}$, where the extra factor $m2^{-n}$ is essentially the probability of one hit and $m-1$ misses during $m$ independent Bernoulli trial with the hit probability $2^{-n}$ (if we regard the event of entering line \ref{step:repeat} as a hit). Carrying this argument to the general case if the algorithm enters line \ref{step:repeat} $k$ times, then we need to consider all possible ways of distributing the $k$ hits onto the $m$ iterations on line \ref{step:allins}. This gives
\begin{equation}
\text{Pr}(G|\text{Algorithm \ref{alg:dummyfree} entered line \ref{step:repeat} in total $k$ times})=\sum_{(k_1,\cdots,k_m)}2^{-mn}\cdot\begin{pmatrix}k \\ k_1,k_2,\cdots,k_m\end{pmatrix}\cdot(2^{-n})^{k_1+k_2+\cdots+k_m}
\end{equation}
where the summation is over the set of non-negative integers $k_1$ through $k_m$ that sums up to $k$. Then Equation \ref{eq:prG} leads to
\begin{equation}
\begin{array}{ccl}
\text{Pr}(G) & = & \displaystyle 2^{-mn}\cdot\sum_{k=0}^\infty\sum_{(k_1,\cdots,k_m)}2^{-kn}\begin{pmatrix}k \\ k_1,\cdots,k_m\end{pmatrix} \\[0.1in]
& = & 2^{-mn}(1+2^{-n}+2^{-2n}+\cdots)^m \\[0.1in]
& = & \displaystyle 2^{-mn}\left(\frac{1}{1-2^{-n}}\right)^m \\[0.1in]
& = & (2^n-1)^{-m}.
\end{array}
\end{equation}
\end{proof}

\end{document}